\documentclass[a4paper]{article}

\usepackage{INTERSPEECH2020}
\usepackage{tabularx}
\title{Dual-Signal Transformation LSTM Network for Real-Time Noise Suppression}
\name{Nils L. Westhausen$^{1}$ and Bernd T. Meyer$^{1}$}
\address{
  $^1$Communication Acoustics \& Cluster of Excellence Hearing4all\\
  Carl von Ossietzky University, Oldenburg, Germany}
\email{nils.westhausen@uol.de, bernd.meyer@uol.de}

\begin{document}

\maketitle

\begin{abstract}
  This paper introduces a dual-signal transformation LSTM network (DTLN) for real-time speech enhancement as part of the Deep Noise Suppression Challenge (DNS-Challenge). This approach combines a short-time Fourier transform (STFT) and a learned analysis and synthesis basis in a stacked-network approach with less than one million parameters. The model was trained on 500\,h of noisy speech provided by the challenge organizers. The network is capable of real-time processing (one frame in, one frame out) and reaches competitive results.
  Combining these two types of signal transformations enables the DTLN to robustly extract information from magnitude spectra and incorporate phase information from the learned feature basis. The method shows state-of-the-art performance and outperforms the DNS-Challenge baseline by 0.24 points absolute in terms of the mean opinion score (MOS). 
\end{abstract}

\section{Introduction}
The task of noise suppression is an important discipline in field of speech enhancement; it is for instance of special importance in work-from-home scenarios where a robust and effective noise reduction can improve the communication quality and thereby reduce the cognitive effort of video conferencing.  
With the uprising of deep neural networks, several novel approaches for audio processing methods based on deep models were proposed \cite{xu2013experimental, han2014learning,pascual2017segan, park2016fully}. 
However, these have often been developed for offline processing which does not require real-time capabilities or the consideration of causality in the processing chain. 
Such models process complete sequences and exploit past and future information of the signals to suppress undesired signal parts. 
Classic signal processing algorithms \cite{ephraim1984speech, griffiths1982alternative} often work on sample or frame level to provide a low input-output delay. 
When designing frame-based algorithms with neural networks, recurrent neural networks (RNN) are a common choice. 
RNNs have produced convincing results in the field of speech enhancement  \cite{valentini2016investigating,valin2018hybrid} and speech separation \cite{isik2016single,kolbaek2017multitalker,luo2018tasnet}. 
Long short term memory networks (LSTM) \cite{hochreiter1997long} represent the state-of-the-art in separation \cite{luo2019dual}. The best-performing networks are often build in a non-causal way by using bidirectional LSTMs where the time sequence is processed causally as well in the reversed direction. 
Bidirectional RNNs always require a full sequence as input and are therefore principally not suited for real-time frame processing. 

The baseline system of the the deep-noise-suppression challenge (DNS-Challenge) \cite{DNSChallenge2020} is called NSNet \cite{9054254} and is also based on RNN layers and provides real-time capability by calculating one output frame per input frame. 
Based on the log power spectrum of the short-time Fourier transform (STFT) of the noisy time signal, this model predicts a gain or mask which is applied to a noisy STFT. 
The predicted speech signal is reconstructed by using the estimated magnitude and the phase of the noisy mixture. 
This approach results in a competitive baseline system, but it does not incorporate any phase information, which could be useful for enhanced speech quality.
Different approaches are tackling phase estimation such as estimating the masks for the real and imaginary part of the STFT instead of the magnitude \cite{williamson2015complex} or calculating an iterative phase reconstruction \cite{wang2018end}. 
Research studies such as \cite{luo2018tasnet, luo2019conv, maciejewski2020whamr}  have shown promising results for speaker separation tasks with a learned analysis and synthesis basis that is not decoupling magnitude and phase information. 
The representation is calculated by multiplying time-domain frames with learned basis functions. This approach was also applied in \cite{kavalerov2019universal} for  separating speech and noise. 

The motivation of the current study is to merge both analysis and synthesis approaches in one model by using a stacked dual signal transformation LSTM network (DTLN). Stacked or cascaded networks were already used in the Deep Clustering speaker separation approach \cite{isik2016single} where an additional enhancement network was added after the separation network. 
In related research, cascaded models were used for denoising and dereverberation \cite{maciejewski2020whamr}. 
The proposed model presented here cascades two separation cores, the first features an STFT signal transformation while the second used a learned signal representation similar to \cite{luo2019conv}. This order was chosen to create a robust magnitude estimation with the first core and enable the second core to further enhance the signal with phase information.
This combination is explored for the first time in the context of noise reduction and could provide beneficial effects due to the complementarity of classic and learned features transformations while maintaining a relatively small computational footprint. 
The stacked network in this paper is considerably smaller as most previously proposed LSTM networks and ensures real-time capability in terms of computational complexity. 

\section{Methods}

\subsection{Signal transformations}
In speaker separation, a time-frequency masking approach is often chosen to separate the speakers' signals. 
Noise suppression is a related source separation problem, but is different in that it only returns the speech signal and discards the noise. 
In the time frequency domain, the separation problem can be formulated as follows:
The microphone signal $y$ is described by
\begin{equation}
    y[n] = x_s + x_n
\end{equation}
where $ x_s $ and $x_n$ are the speech and noise components of time signals, respectively. 

In a noise suppression task, the desired signal is the speech signal. When the signal $y$ is transformed with an STFT in a complex time-frequency representation (TF), the TF representation of the estimated speech signal $\hat{X}_s$ can be predicted as follows:
\begin{equation}
    \hat{X}_s(t,f) = M(t,f) \cdot  |Y(t,f)| \cdot e^{j \phi y},
\end{equation}
where $|Y|$ is the magnitude of the STFT of $y$. $M$ is a mask (with masking values ranging from 0 to 1) that is applied to $Y$, and $e^{j \phi y}$ is the phase of the noisy signal. $\hat{X}_s$ can now be transformed back with an inverse STFT to $\hat{x}_s$. 
In this formulation, the phase of the noisy signal is used to predict the clean speech signal. 

The second signal transformation of the DTLN was first proposed by Luo and colleagues \cite{luo2018tasnet}. 
The formulation of the approach is described in the following: The mixture is split into overlapping frames $y_k$ of length $L$ with frame index $k$. 
The frames are multiplied by $U$, which has $N \times L$ learned basis functions
\begin{equation}
    w_k = y_k U
\end{equation}
to create the feature representation $w_k$ with dimension $N \times 1$ of frame $y_k$. To recover the speech representation $d_k$ from $w_k$, a mask $m_k$ can be estimated given by
\begin{equation}
    \hat{d}_k = m_k \cdot w_k,
\end{equation}
where $\hat{d}_k$ is the feature representation at index $k$ of the estimated speech signal. $\hat{d}_k$ can be transformed back to the time domain by
\begin{equation}
    \hat{x}_k = \hat{d}_k V,
\end{equation}
where $V$ contains $N$ learned basis functions of length $L$. $\hat{x_k}$ is the estimated frame at index $k$.
The estimated time signal $\hat{x}_s$ is reconstructed by using an overlap-add procedure.

\subsection{Network architecture}
\begin{figure}[t]
  \centering
  \includegraphics[width=115pt]{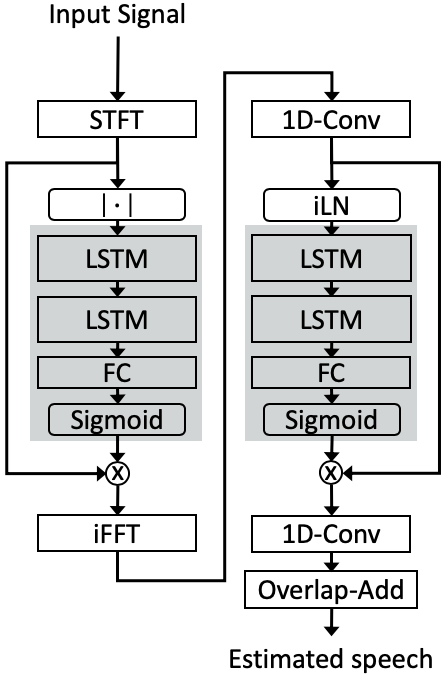}
  \caption{Illustration of the proposed network architecture. The processing chain on the left shows the first separation core using the STFT signal transformation while the building blocks on the right represent the second core with learned feature transformations based on 1D-Conv layers.}
\label{fig:dns-network}
\end{figure}

The stacked dual-signal transformation LSTM network architecture introduced in this paper has two separation cores containing two LSTM layers followed by a fully-connected (FC) layer and a sigmoid activation to create a mask output. 
The first separation core uses an STFT analysis and synthesis base. 
The mask predicted by the FC layer and the sigmoid activation is multiplied by the magnitude of the mixture and transformed back to the time domain using the phase of the input mixture, but without reconstructing the waveform. 
The frames coming from the first network are processed by an 1D-Conv layer to create the feature representation. The feature representation is processed by a normalization layer before it is fed to the second separation core. 
The predicted mask of the second core is multiplied with the unnormalized version of the feature representation. The result is used as input to a 1D-Conv layer for transforming the estimated representation back to the time domain. In a last step, the signal is reconstructed with an overlap and add procedure. The architecture is visualized in Figure~\ref{fig:dns-network}.

To account for the real-time character of the model, instant layer normalization (iLN) is used. 
Instant layer normalization is similar to standard layer normalization \cite{ba2016layer} and was introduced as channel-wise layer normalization in \cite{luo2018conv2}. All frames are normalized individually without accumulating statistics over time and are scaled with the same learnable parameters. 
In the current work, this normalization scheme is referred to as instant layer normalization to differentiate from cumulative layer normalisation \cite{luo2019conv}.
\subsection{Datasets}
The training dataset was created from the provided audio data of the DNS-Challenge. The speech data is part of the Librispeech corpus \cite{panayotov2015librispeech}, and the noise signals originated from the Audioset corpus \cite{gemmeke2017audio}, Freesound and DEMAND corpus \cite{thiemann2013diverse}. 
500\,h of data were created by using the provided scripts. 
The default SNR range (0 to 40\,dB) was changed to -5 to 25~dB to include negative SNRs  and limit the total range. To cover a more fine-grained SNR distribution, the number of SNR levels was increased from 5 to 30. All further parameters remained unchanged. 
The 500\,h dataset was divided into training (400\,h) and cross validation data (100\,h), which corresponds to the common 80:20~\% split. 
All training data was sampled at~16~kHz.

The challenge organizers also provided a test set which contains four different categories each containing 300 samples. The categories are synthetic clips without reverb, synthetic clips with reverb, real recordings collected internally at Microsoft and real recordings from Audioset. 
The synthetic data was taken from the Graz University’s clean speech dataset \cite{pirker2011pitch}.
The SNRs of the synthetic data were randomly distributed from 0 to 25 dB SNR. 
The impulse responses of the reverberant data were measured in multiple rooms at Microsoft with reverberation times (RT60) ranging from 300 to 1300 ms. 
Further, a blind test set was created by the organizers which is evaluated in an ITU P-808 \cite{recommendation2018itu} setup. 
The full details of training and test sets are provided in \cite{DNSChallenge2020}.

To correctly estimate the performance with all objective measures in a noisy reverberant environment, the reverberant single speaker and noise test set of the WHAMR corpus \cite{maciejewski2020whamr} at 16~kHz sampling frequency was used. 
We turned to this dataset because some objective measures need a properly delayed but clean reference signal for correct calculation. 
Since these signals are not provided in the DNS-Challenge test set, we used the WHAMR dataset, which has clean non-reverberant speech files accounting for the delay of the impulse response.
The used WHAMR test set consists of 3000 mixtures. The speech files are taken from the WSJ0-mix corpus \cite{Hershey2016mar} that are often used in speaker separation. 
The speech files are convolved with room impulse responses with the RT60 ranging from 100 to 1000\,ms that were simulated with Pyroomacoustics \cite{scheibler2018pyroomacoustics}. 
The noise consists of real-life recordings of situations such as coffee shops, restaurants, bars, office buildings and parks. The SNRs range from -3 to 6 dB relative to the speech.

\subsection{Model configuration and training setup}
The DTLN in this paper\footnote{A Keras implementation of the DTLN can be found at https://github.com/breizhn/DTLN} has 128 units in each of its four LSTM layers. The frame size is 32\,ms and the shift 8\,ms. The FFT size is 512 and equal to the frame length. The 1D-Conv Layer to create the learned feature representation has 256 filters. During training, 25\% of dropout is applied between the LSTM layers. The Adam optimizer is used with a learning rate of 1e-3 and and a gradient norm clipping of 3. The learning rate is halved if the loss on the validation set does not improve for three consecutive epochs. Early stopping is applied if loss on the validation set does not decrease for ten epochs. The model is trained on a batch size of 32, and each sample has the length of 15\,s. The average time for one training epoch on a Nvidia RTX 2080 TI is around 21 minutes. 

As training objective the scale-sensitive negative SNR \cite{kavalerov2019universal} was used.
Compared to the Scale Invariant Signal to Noise Ratio (SI-SNR) \cite{luo2018tasnet} it should avoid possible level offsets between input mixture and predicted cleaned speech, which is desirable in real-time-processing systems.
Also, since it operates in time domain, the phase information can implicitly be considered. In contrast the Mean Squared Error between the estimated and clean magnitude STFT of the speech signal as training objective is not able to use any phase information in the optimization process.

\subsection{Baselines}
The first baseline is the noise suppression network (NSNet) provided by the challenge organizers. NSNet was optimized with an MSE-based speech distortion weighted loss in frequency domain and was trained on a rather small corpus of 84\,h of speech and noise mixtures. It consists of three recurrent layer with 256 gated recurrent units (GRU) \cite{chung2014gru} and a fully-connected layer with sigmoid activation for mask prediction. The frame size is 20\,ms and the frame shift\,10 ms. GRUs are similar to LSTMs but without a cell state passed over time. 

Additionally, our DTLN method is compared to four models with the same training setup as the proposed model: 
These models are explored to quantify the effect of using one feature representation only in two different topologies (stacked versus densely-connected LSTMs): 
The first and the second model consist of four LSTM layers followed by a fully-connected layer with a sigmoid activation to predict a mask. 
The first one (B1) uses an STFT analysis and synthesis basis, while B2 used a learned basis of size 256. The third (B3) and the fourth model (B4) are stacked models similar to the proposed method. Both separation kernels of B3 are using an STFT base. B4 has a learned feature base of size 256 for both separation kernels. 
The size of the LSTM layers is chosen with the aim of obtaining a similar size as the DTLN method in terms of the number of parameters. 
The configurations are again shown in Table~\ref{tab:models}. \\
\begin{table}[th]
  \caption{Number of parameters and RNN units in each layer for the proposed DTLN approach as well as for the baseline systems.}
  \label{tab:models}
  \centering
  \begin{tabular}{ l r r}
    \toprule
    \textbf{Method} & \textbf{\# Prams} & \textbf{\# Units}\\
    \midrule
     NSNet (3 Layer, STFT) &   1.27 M    &     256     \\
     \midrule
     B1 (4 Layer, STFT)    &   988 K     &     166    \\
     B2 (4 Layer, learned) &   984 K     &     139    \\
     B3 (2x2 Layer, STFT)  &   988 K     &     156    \\
     B4 (2x2 Layer, learned) &   987 K     &     95   \\
     \midrule
     \textbf{DTLN} (2x2 layer, STFT+learned)  &   987 K     &     128     \\
    
    \bottomrule
  \end{tabular}
  
\end{table}
\subsection{Objective and subjective evaluation}
For comparison of the DTLN approach and the baselines, we use three objective measures, i.e., the Perceptual Evaluation of Speech Quality (PESQ) \cite{pesq}, the Scale-Invariant Signal to Distortion Ratio (SI-SDR) \cite{le2019sdr} and the Short Time Objective Intelligibility measure (STOI) \cite{taal2010short}.

The subjective evaluation was performed with the a ITU-T P.808 setup on the Amazon mechanical Turk (AMT) implemented and organized by Microsoft. In total, there were two evaluation runs, one on the known test data set of the DNS-Challenge and one on a blind test set provided later on. Each file was rated by five or ten judges in the first and second run, respectively. 
\section{Results}
\begin{table*}[t]
    \caption{Results in terms of PESQ [MOS], SI-SDR [dB] and STOI [\%] of the non reverberant test set, the reverberant test set of the DNS challenge and the reverberant single mix test set of the WHAMR corpus.}
  \centering
  \begin{tabular}{l r r r r r r r r r r r r}
  \toprule
  &  \multicolumn{3}{c}{\textbf{DNS test set no reverb}} & & \multicolumn{3}{c}{\textbf{DNS test set with reverb}} &  & \multicolumn{3}{c}{\textbf{WHAMR test set}} 
        \\
    \midrule
    \textbf{Method} & \textbf{PESQ} &  \textbf{SI-SDR}  & \textbf{STOI} &
    & \textbf{PESQ} & \textbf{SI-SDR}  & \textbf{STOI} &
    & \textbf{PESQ} & \textbf{SI-SDR}  & \textbf{STOI}\\
    \midrule
     Noisy &  2.45 & 9.07 & 91.52 &  & 2.75 & 9.03 & \textbf{86.62} & & 1.83 &	-2.73	&	73.00 \\
     \midrule
     NSNet  & 2.70 & 12.47 & 90.56 & & 2.47 & 9.18 & 82.15 & & 1.91	&	0.34	&	73.02 \\
     B1     & 3.00 & 16.05 & 94.53 & & 2.75 & \textbf{11.33} & 85.41 & & 2.20	&	1.95	&	79.93 \\
     B2     & 3.00 & 15.87 & 94.22 & & 2.74 & 10.92 & 85.05 & & 2.18 &	1.88	&	79.34 \\
     B3     & 3.03 & 16.27 & 94.74 & & 2.70 & 10.84 & 84.80 & & \textbf{2.23}	 &	1.94	&	80.23	 \\
     B4     & 2.96 & 15.51 & 93.86 & & \textbf{2.76} & 10.77 & 84.90 & & 2.20	& 1.80    &	78.90	 \\
     \midrule
     DTLN   & \textbf{3.04} & \textbf{16.34} & \textbf{94.76} & & 2.70 & 10.53 & 84.68 & & \textbf{2.23}	&	\textbf{2.12}	&	\textbf{80.40}	\\
     \bottomrule
  \end{tabular}
  \label{tab:results}
\end{table*}
The results of the objective evaluation are shown in Table~\ref{tab:results} and the subjective evaluation in Table~\ref{tab:mos_ver1}. The results are described in the following:
\\
\textbf{Objective results for the non-reverberant DNS-Challenge test set}: In the non-reverberant condition, all models produce improvements over the noisy condition. 
NSNet is outperformed 
by the DTLN and all additional baselines. 
All models trained on 500\,h of data are producing similar results. The best results in terms of PESQ, SI-SDR and STOI were reached by the DTLN network. 
The high values obtained with B3 and the DTLN show the strength of stacked models. Even though B4 is also a stacked model, it performed considerably worse, which is discussed in Section\,4. 
\\
\textbf{Objective results for the reverberant DNS-Challenge test set}: In this condition, results are not as clear as in the non-reverberant condition. 
In terms of PESQ, only B4 shows a slight improvement over the noisy condition. 
For the SI-SDR all models show an improvement, while STOI predicts the highest quality for the original noisy condition. 
One issue with the intrusive or double-ended measures is that they require a reference signal which is in this case the reverberant clean speech.
With this reference signal a potential dereverberation effect by any speech enhancement model would result in a decrease of the objective measures, which presumably is an important factor for these results. 
\\
\textbf{Objective results on the WHAMR test set}: All methods are showing an improvement over the noisy condition, with the best scores obtained by the DTLN approach. 
Similar performance levels are again reached by B3. The baseline shows just a slight improvement for all objective measures. It should be mentioned that the mixtures used in this corpus have a smaller SNR range around 0 and is therefore a more challenging condition for the models. 
\\
\textbf{Subjective results on the DNS-Challenge test sets}:
The subjective results for the known non-reverberant test set are in line with the objective results. 
For the reverberant test set, the subjective evaluation shows a clear benefit for DTLN relative to the noisy condition and the baseline. This effect is not reflectFor ed by the objective measures with exception of the SI-SDR, which shows some improvement over the baseline and the noisy condition. 
The decrease in quality of the NSNet predicted by PESQ and STOI in the reverberant condition is also observed in the subjective data. 
Consistent results are obtained with the real recordings, both for the conditions \emph{known} and \emph{blind}.
\begin{table}[th]
  
  \caption{Subjective ratings in terms of the MOS for the known and blind test set of the DNS-Challenge. The  overall 95\% confidence intervals for the known and blind test sets are 0.04 and 0.02, respectively.}
  \centering
  \begin{tabularx}{0.46\textwidth}{ X X X X X X X}
    \toprule
    \textbf{Method} &  \multicolumn{2}{c}{\textbf{No}} & \multicolumn{2}{c}{\textbf{With}}   & \multicolumn{2}{c}{\textbf{Real}} 
        \\
       &  \multicolumn{2}{c}{\textbf{reverb}} & \multicolumn{2}{c}{\textbf{reverb}}   & \multicolumn{2}{c}{\textbf{recordings}} \\
       \midrule
     & known & blind	&	known & blind	&	known & blind	 \\
    \midrule
     Noisy & 3.02 & 3.32	&	2.44 & 2.78 	&	3.01 & 2.97		 \\ 
     NSNet & 3.14 & 3.49	&	2.16 & 2.64	&	2.99 & 3.00		 \\ 
     
     DTLN & \textbf{3.41} & \textbf{3.58}	&	\textbf{2.56} & \textbf{2.95}	&	\textbf{3.14} & \textbf{3.21}		\\  
        
    \bottomrule
  \end{tabularx}
  \label{tab:mos_ver1}
\end{table}\\
\textbf{Results on execution time}: In the context of the DNS-Challenge, the execution time of one 32\,ms frame on a quad-core I5 6600K CPU was measured. The measurement was performed by either processing a complete sequence or by using frame-wise processing. 
Execution times of 0.23\,ms and 2.08\,ms were measured for the sequence and frame-wise processing, respectively. 
The large difference between sequence and frame processing can be explained by the overhead caused by calling models for prediction in Keras. Converting the model to Tensorflow's SavedModel format reduces the execution time for frame by frame processing to 0.65\,ms, which is a large improvement. However, the sequence processing time is nearly three times lower and demonstrates the potential performance on a CPU.
\section{Discussion}
In the following, we first discuss differences between baseline systems, which also has implications for the components of the DTLN system. 
The results on the non-reverberant, the reverberant and the WHAMR test are showing better results for the systems B1 and B3 (using STFT features) than for B2 and B4 (that used learned feature representations). 
One potential reason for the better performance with STFTs is the fixed number of parameters across networks, and - since the STFT is fixed and rule-based - it is possible that B1 and B3 exploit the higher number of parameters available for LSTM layers in comparison to learned-feature approaches. 

Secondly, we assume that STFT features provide a higher robustness for noisy input since phase information - which is not useful in high-noise conditions - is discarded. 
Vice versa, networks using learned features have to determine masks implicitly for both magnitude and phase information. 
Another possible reason for the difference could be the compression which is performed by the learned feature representation in this work. The learned feature representation maps 512 audio samples to a feature representation of size 256. Feature representations with greater size would have cost even more parameters and it was empirically found that the reduction of the feature representation doesn't have a great impact on the speech quality of the proposed model. 

The results also show that stacking networks using STFT and learned feature transformation slightly improves overall baseline systems by using fewer LSTM units than the pure STFT systems. 
LSTM units are computational more complex as fully-connected or 1D-Conv layers, i.e., a reduction of units is especially desirable for this network type.
However, the relatively small difference in terms of objective measures between DTLN and the related systems (B1-B4) also suggests that a part of the performance is generated by the large amount of training data and the training setup.

\section{Conclusions}
This paper introduced an approach for noise suppression based on a stacked dual signal transformation LSTM network for real-time speech enhancement, which was trained on a large-scale data set. We were able to show an advantage of using two types of analysis and synthesis bases in a stacked network approach. The DTLN model works robustly in noisy reverberant environments.
Although we combined a basic training setup with a straight-forward architecture, we observed absolute improvements of 0.22 in terms of MOS over all subjective evaluations relative to the noisy conditions. 

\section{Acknowledgements}
This research was supported by the DFG (Cluster of Excellence 1077/1 Hearing4all; URL: http://hearing4all.eu). The architecture was partially developed on a GPU donated by the Nvidia GPU Grant program. Thanks goes to challenge organizers from Microsoft for conducting the DNS-Challenge and providing the data and the scripts.

\bibliographystyle{IEEEtran}

\bibliography{mybib}

\end{document}